%% file: spca.tex
\documentclass[conference]{IEEEtran}
\IEEEoverridecommandlockouts

\usepackage{cite}

\include{commands}

\usepackage{commath}

\newcommand{\EE}{\mathbf{E}}

\DeclareMathOperator*{\argmin}{arg\,min}
\DeclareMathOperator{\prox}{prox}
\DeclareMathOperator{\sgn}{sgn}

\global\long\def\b{\beta}

\global\long\def\w{\omega}

\begin{document}
%
\title{Online Learning for Sparse PCA in High Dimensions: Exact Dynamics and Phase Transitions}

\author{\IEEEauthorblockN{Chuang Wang and Yue M. Lu}\vspace{1ex}
\IEEEauthorblockA{John A. Paulson School of Engineering and Applied Sciences\\
Harvard University}
\thanks{This work was supported in part by the NSF under grant CCF-1319140 and by ARO under grant W911NF-16-1-0265.}
}


\maketitle

\begin{abstract}
We study the dynamics of an online algorithm for learning a sparse leading eigenvector from samples generated from a spiked covariance model. This algorithm combines the classical Oja's method for online PCA with an element-wise nonlinearity at each iteration to promote sparsity. In the high-dimensional limit, the joint empirical measure of the underlying sparse eigenvector and its estimate provided by the algorithm is shown to converge weakly to a deterministic, measure-valued process. This scaling limit is characterized as the unique solution of a nonlinear PDE, and it provides exact information regarding the asymptotic performance of the algorithm. For example, performance metrics such as the cosine similarity and the misclassification rate in sparse support recovery can be obtained by examining the limiting dynamics. A steady-state analysis of the nonlinear PDE also reveals an interesting phase transition phenomenon. Although our analysis is asymptotic in nature, numerical simulations show that the theoretical predictions are accurate for moderate signal dimensions.
\end{abstract}


\section{Introduction}

Consider the spiked covariance model \cite{Johnstone:2001}, where we are given a sequence of $p$-dimensional sample vectors $\vy_1, \vy_2, \ldots$ that are distributed according to
\begin{equation}\label{eq:spiked}
\vy_k = \sqrt{\frac{\omega}{p}} c_k \vxi + \va_k.
\end{equation}
Here, $\vxi$ is an unknown vector in $\R^p$, $c_k \sim \mathcal{N}(0, 1), \va_k \sim \mathcal{N}(0, \mI_p)$, and $\omega$ is a positive quantity specifying the signal-to-noise ratio (SNR); $(c_i, \va_i)$ and $(c_j, \va_j)$ are independent for $i \neq j$. In this paper, we analyze the exact dynamics of an online (incremental) algorithm for estimating $\vxi$ in the high-dimensional ($p \rightarrow \infty$) limit.

The model in \eref{spiked} arises in the theoretical study of principal component analysis (PCA), an important statistical tool in exploratory data analysis, visualization and dimension reduction. A standard method to estimate $\vxi$ is to compute the leading eigenvector of the sample covariance matrix $\mat{\Sigma} = \frac{1}{n} \sum_{k=1}^n \vy_k \vy_k^T$. For fixed $p$ and when the number of samples $n$ tends to infinity, the eigenvector is a consistent estimator of $\vxi$ (up to a normalization constant.) However, in the regime where $p$ and $n$ are both large and comparable in size, the estimate given by the eigenvector is no longer consistent \cite{Nadler:2008,Johnstone:2009}.

To address this issue, a flurry of work---under the name of sparse PCA---has exploited the sparsity structure of $\vxi$ (see, \emph{e.g.},\cite{Jolliffe:2003, Zou:2006, Johnstone:2009}.) In addition to potentially improving the estimate of $\vxi$, sparse PCA generates a more parsimonious and interpretable representation, using a small subset of feature variables to explain the original data.

The natural formulation of sparse PCA leads to nonconvex optimization problems \cite{Jolliffe:2003, Zou:2006, Johnstone:2009}. Convex relaxations via semidefinite programming (SDP) \cite{dAspremont:2007, Amini:2009} are possible, but the computational and storage cost of SDP may become prohibitive when the dimensionality is high. Many efficient algorithms have been proposed to solve sparse PCA, in both offline \cite{Johnstone:2009, Shen:2008, Journee:2008, Ma:2013, Deshpande:2014} and online \cite{Mairal:2009, Arora:2012, Balsubramani:2013, Yang:2015} settings. In the latter case, which is the setting we study in this paper, sample vectors $\set{\vy_k}$ arrive sequentially in an infinite stream; as soon as a new sample vector (or a small batch of them) has arrived, an online algorithm computes an instantaneous update to its estimate of $\vxi$. Since they only keep and operate on small sets of current samples, online algorithms are memory and computationally efficient. Moreover, as they provide estimates \emph{on-the-fly}, online algorithms are well-suited to dynamic scenarios where the principal component vectors can be time-varying.


In this paper, we analyze an online sparse PCA algorithm that combines the classical Oja's method \cite{Oja:1985} with an element-wise nonlinearity (\emph{e.g.}, soft-thresholding) at each iteration to promote sparsity (see \sref{algorithm} for the exact form.) Specifically, let $\vx_k$ be the estimate of $\vxi$ given by the algorithm upon receiving the $k$th sample; let $x_k^i$ and $\xi^i$ denote the $i$th component of each vector. Also, define the joint \emph{empirical} measure of $\vx_k$ and $\vxi$ as
\begin{equation}\label{eq:mu}
\mu_k^p(x, \xi) \bydef \frac{1}{p} \sum_{i= 1}^p \delta(x - x_k^i, \xi - \xi^i).
\end{equation}
Note that $\mu_k^p(x, \xi)$ is a random element in $\mathcal{M}(\R^2)$, the space of probability measures on $\R^2$. As the main result of this work, we show that, as $p \rightarrow \infty$ and with suitable time-rescaling, the sequence of empirical measures $\set{\mu_k^p(x, \xi)}_p$ converges weakly to a deterministic measure-valued process $\mu_t(x, \xi)$. Moreover, this limiting measure $\mu_t(x, \xi)$ is the unique solution of a nonlinear partial differential equation (PDE.)

The deterministic \emph{scaling limit} as specified by the PDE and its solution provides a wealth of information regarding the performance of the online sparse PCA algorithm. For example, the limiting value of the cosine similarity
\begin{equation}\label{eq:q}
Q^p_k \bydef \frac{\vx_k^T \vxi}{\norm{\vx_k}\norm{\vxi}}
\end{equation}
at any step $k$ can be easily obtained by computing the expectation $\EE(x\xi)$ with respect to the limiting measure $\mu_t(x, \xi)$. More involved questions, such as the misclassification rate in sparse support recovery, can also be answered by examining $\mu_t(x, \xi)$. Finally, studying the PDE in its steady-state leads to an exact characterization of the long-time behavior of the online sparse PCA algorithms. This steady-state analysis also uncovers a phase transition phenomenon: the performance of the algorithm can exhibit markedly different behaviors depending on the parameter settings and SNR values.

The rest of the paper is organized as follows. In \sref{algorithm}, we give the details of the online sparse PCA algorithm that we analyze in this work. The scaling limit of the algorithm is presented in \sref{limit}. As a special case, we study in \sref{Oja} the classical Oja's method and derive an analytical expression characterizing the limiting cosine similarity between its estimates and $\vxi$. Finally, a steady-state analysis and an associated phase transition phenomenon are discussed in \sref{steady}.

\section{Online Algorithm for Sparse PCA }
\label{sec:algorithm}

We consider the online setting, where sample vectors $\set{\vy_k}$ arrive sequentially. We assume that the samples are generated by the spiked covariance model in \eref{spiked} with a single leading eigenvector $\vxi$. We further assume that each element of $\vxi$ is an i.i.d. sample drawn from a mixture distribution
\begin{equation}\label{eq:prior}
\pi(\xi) = (1-\rho) \delta(\xi) + \rho \, u(\xi),
\end{equation}
where $\rho \in (0, 1]$ is a parameter controlling the sparsity level, and $u(\xi)$ is a density function such that $\int \xi^2 u(\xi) \dif \xi = 1/\rho$. The preceding requirement makes sure that ${\norm{\vxi}}/{\sqrt{p}} \rightarrow 1$ as the dimension $p \rightarrow \infty$. An example of \eref{prior} is the standard Bernoulli-Gaussian distribution. By choosing
\[
u(\xi) = [\delta(\xi - 1/\sqrt{\rho}) + \delta(\xi + 1/\sqrt{\rho})]/2,
\]
the distribution in \eref{prior} can also describe the sparse signal model considered in \cite{Amini:2009}.

In this work, we analyze a simple recursive algorithm for estimating $\vxi$ from the stream of samples $\set{\vy_k}$. The algorithm starts from some initial estimate $\vx_0$. Upon receiving the $k$th data sample $\vy_k$, it updates its estimate as follows:
\begin{equation}\label{eq:oist}
\begin{aligned}
\widetilde{\vx}_{k} & = \vx_{k-1}+ ({\tau}/{p}) \, \vy_k \vy_k^T \vx_{k-1}\\
	\vx_{k}          & = \sqrt{p} \, \eta(\widetilde{\vx}_k) / \norm{\eta(\widetilde{\vx}_k)}.
\end{aligned}
\end{equation}
Here, $\tau > 0$ is the step size, and $\eta(\cdot)$ is an element-wise nonlinear mapping taking the form
\begin{equation}\label{eq:eta}
\eta(x) = x - \frac{1}{p} \phi(x),
\end{equation}
for some piecewise smooth function $\phi: \R \rightarrow \R$. Clearly, the method is online (incremental): it processes one sample at a time. Once a sample has been processed, it will be discarded and never used again. 

The update steps in \eref{oist} as well as the expression in \eref{eta} need some explanations. First, we note that, without the nonlinear mapping (\emph{i.e.}, by setting $\eta(x) = x$), the recursions in \eref{oist} are exactly the original Oja's method \cite{Oja:1985} for online PCA. The nonlinearity \eref{eta} in $\eta(\cdot)$ is introduced to promote sparsity of the estimates. To see this, we consider an optimization formulation for sparse PCA in the \emph{offline} setting:
\begin{equation}\label{eq:offline}
\widehat{\vx} = \underset{\norm{\vx} = \sqrt{p}}{\argmin} \ \frac{-\vx^T \mat{\Sigma} \vx}{2} + \sum_{i=1}^p \Phi(x^i),
\end{equation}
where $\mat{\Sigma}$ is the population (or sample) covariance matrix, and $\Phi(\cdot)$ is an element-wise penalty function that favors sparse solutions. For example, $\Phi(x) = \lambda \abs{x}$  for lasso-type penalizations; or we can choose $\Phi(x) = \lambda_1 x^2 + \lambda_2 \abs{x}$ for the elastic net \cite{Zou:2006}. To solve \eref{offline}, we use a proximal gradient method \cite{Parikh:2014} followed by a projection onto the sphere of radius $\!\sqrt{p}$:
\[
\begin{aligned}
\widetilde{\vx}_k &= \vx_{k-1} + (\tau/p) \mat{\Sigma}\vx_{k-1}\\
\vx_k &= \sqrt{p} \; \prox_{{\tau}\Phi/p} (\widetilde{\vx}_k) \norm{\prox_{{\tau}\Phi/p} (\widetilde{\vx}_k)}^{-1},
\end{aligned}
\]
where $\prox_{{\tau}\Phi/p}$ denotes the proximal operator of the function $\tau \Phi(x) / p$. Replacing the covariance matrix $\mat{\Sigma}$ by its instantaneous (and noisy) version $\vy_k \vy_k^T$ and using the approximation $\prox_{{\tau}\Phi/p}(x) \approx x - \tau (\tpd{}{x}\!\Phi)/p$ (see, \emph{e.g.}, \cite[p. 138]{Parikh:2014} for a justification of this approximation which holds for large $p$), we reach our algorithm in \eref{oist} as well as the form given in \eref{eta}.

\begin{example}\label{ex:oist}
Consider a lasso-type penalization in \eref{offline} where $\Phi(x) = \frac{\beta}{\tau} \abs{x}$ for some $\beta > 0$. The associated proximal operator is the standard soft-thresholding function with parameter $\beta/p$, which can be approximated, for large $p$, as
\[
\prox_{{\tau}\Phi/p}(x) \approx x - \frac{\beta \sgn(x)}{p}.
\]
This corresponds to choosing $\phi(x) = \beta \sgn(x)$ in \eref{eta}. In what follows, we refer to this particular variant of the algorithm as Oja's algorithm with iterative soft thresholding (OIST for short.)
\end{example}

\section{Dynamics in High Dimensions: Scaling Limits}
\label{sec:limit}

In what follows, we analyze the dynamics of the online sparse PCA algorithm in \eref{oist} in the large $p$ limit. The central object in our analysis is the empirical measure $\mu_k^p(x, \xi)$ as defined in \eref{mu}. Here, the subscript $k$ indicates the iteration step, and the superscript $p$ makes explicit the dependence of the measure on the dimension $p$. 

The measure $\mu_k^p$ contains a great deal of information about the algorithm. For example, using the notation
\[
\inprod{f, \mu_k^p} \bydef \frac{1}{p} \sum_{i \le p} f(x_k^i, \xi^i),
\]
for a test function $f(x, \xi)$, we can write the cosine similarity defined in \eref{q} as $Q^p_k = {\inprod{x\xi, \mu_k^p}}/{\sqrt{\inprod{\xi^2, \mu_k^p}}}$. Similarly, more involved quantities such as the misclassification rate in sparse support recovery can also be written in terms of $\mu_k^p$.

\subsection{The Main Convergence Result}

To establish the scaling limit of $\mu_k^p$, we first embed the discrete-time sequence in continuous-time by defining
\[
\mu_t^p \bydef \mu_{\lfloor pt \rfloor}^p,
\]
where $\lfloor \cdot \rfloor$ is the floor function. Similarly, we can define $Q^p_t$ as the continuous-time rescaled version of $Q^p_k$. Note that this type of time embedding and rescaling is standard in studying the convergence of stochastic processes \cite{Billingsley:1999}. (Some technicalities before we move on: since the empirical measure is random, $\mu_t^p$ is a piecewise-constant c\`{a}dl\`{a}g process taking values in $\mathcal{M}(\R^2)$, the space of probability measures on $\R^2$. In short, $\mu_t^p$ is a random element in $D(\R^+, \mathcal{M}(\R^2))$, for which the notion of weak convergence is well-defined. See, \emph{e.g.}, \cite{Kallenberg:2002}.)

\begin{theorem}\label{thm:limit}
Suppose that $\mu_0^p$, the empirical measure at time $k = 0$, converges (weakly) to a deterministic measure $\mu_0 \in \mathcal{M}(\R^2)$ and that $Q_0 = \inprod{x\xi, \mu_0} \neq 0$. Then, as $p \rightarrow \infty$, the measure-valued stochastic process $\mu_t^p$ converges weakly to a deterministic process $\mu_t$, characterized as the unique solution to the following nonlinear PDE (given in the weak form): for any positive, bounded and $C^3$ test function $f(x, \xi)$,
\begin{equation}\label{eq:weak_form}
\begin{aligned}
\inprod{f, \mu_t} &= \inprod{f, \mu_0} + \int_0^t \inprod{\Gamma\left(x, \xi, Q_s, R_s\right) \tpd{}{x}\!f, \mu_s} \dif s \\
\qquad &+ \frac{\tau^2}{2}\int_0^t \left(1+\omega  Q^2_s\right) \inprod{ \tpd[2]{}{x}\!f, \mu_s} \dif s,
\end{aligned}
\end{equation}
where
\begin{equation}\label{eq:q_r}
Q_t = \iint x\xi \dif\mu_t, \quad R_t \bydef \iint x\phi(x) \dif\mu_t;
\end{equation}
with $\phi(x)$ being the function introduced in \eref{eta}, and 
\begin{equation}\label{eq:Gamma}
\Gamma(x, \xi, Q, R) \bydef \tau \omega Q \xi  - \phi(x) - x\Big[\tau \omega Q^2  - R + \frac{\tau^2}{2}(1 + \omega Q^2)\Big].
\end{equation}
\end{theorem}
\begin{remark}
The deterministic measure-valued process $\mu_t(x, \xi)$ characterizes the exact dynamics of the online sparse PCA algorithm in \eref{oist} in the high-dimensional limit. The nonlinear PDE \eref{weak_form} specifies the time evolution of $\mu_t(x, \xi)$. Note that \eref{weak_form} is presented in the weak form. If the strong, density valued solution exists, then it must satisfy
\begin{equation}\label{eq:strong_form}
\begin{aligned}
\tpd{}{t}\!P_t(x\,\vert\,\xi) &= - \tpd{}{x}\!\left[\Gamma(x, \xi, Q_t, R_t) P_t(x\,\vert\,\xi)\right]\\
&\qquad\quad+\frac{\tau^2(1+\omega Q_t^2)}{2} \tpd[2]{}{x}\!P_t(x\,\vert\,\xi),
\end{aligned}
\end{equation}
where we use $P_t(x\,\vert\,\xi)$ to denote the conditional density of $x$ given $\xi$ at time $t$. The joint density can then be computed as $P_t(x, \xi) = P_t(x\,\vert\,\xi) \pi(\xi)$, where $\pi(\xi)$ is the marginal density defined in \eref{prior}.
\end{remark}
\begin{remark}
For each $\xi$, the PDE \eref{strong_form} resembles a Fokker-Planck equation \cite{Risken:1996} describing the time-evolution of the probability density associated with a particle undergoing a drift-diffusion process in one spatial dimension. There is, however, one important distinction: the PDEs associated with different values of $\xi$ are \emph{coupled} via the quantities $Q_t$ and $R_t$, which themselves depend on the current densities $P_t(x\,\vert\,\xi)$. To see this, we rewrite \eref{q_r} as
\begin{align}
		Q_t & = \EE_{\xi} \left(\xi \int xP_t(x\,\vert\,\xi)\dif x\right)  \label{eq:qt2}\\
		R_t & = \EE_{\xi} \left(\int x \phi(x)P_t(x\,\vert\,\xi)\dif x\right),\label{eq:rt2}
\end{align}
where $\EE_{\xi}$ denotes the expectation with respect to the variable $\xi$ drawn from the prior distribution $\pi(\xi)$.
\end{remark}

\begin{proposition}\label{prop:q}
Under the same assumptions of Theorem~\ref{thm:limit}, the stochastic process $Q^p_t \bydef Q^p_{\lfloor tp \rfloor}$ converges weakly, as $p \rightarrow \infty$, to the deterministic function $Q_t$ defined in \eref{q_r}. 
\end{proposition}
\begin{remark}
We note that $Q_t^p$ describes the time-evolution of the cosine similarity \eref{q} between the estimate given by the algorithm and the unknown vector $\vxi$. This result shows that the dynamics of $Q_t^p$ converges to a deterministic curve $Q_t$, which can be computed from the limiting measure $\mu_t$.
\end{remark}

\begin{figure}[t]
	\centering
	\includegraphics[scale=0.65]{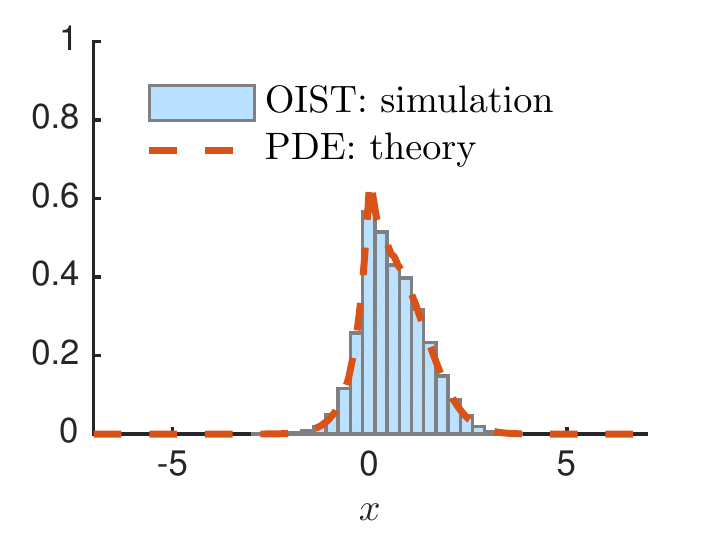}\includegraphics[scale=0.65]{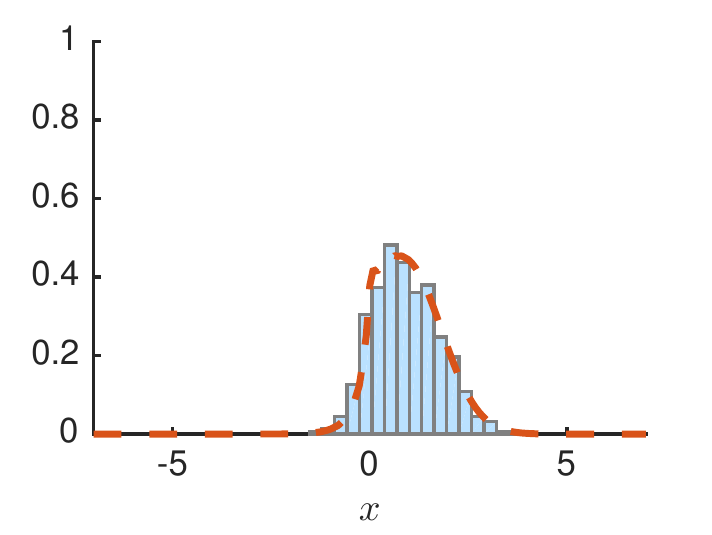}
	\includegraphics[scale=0.65]{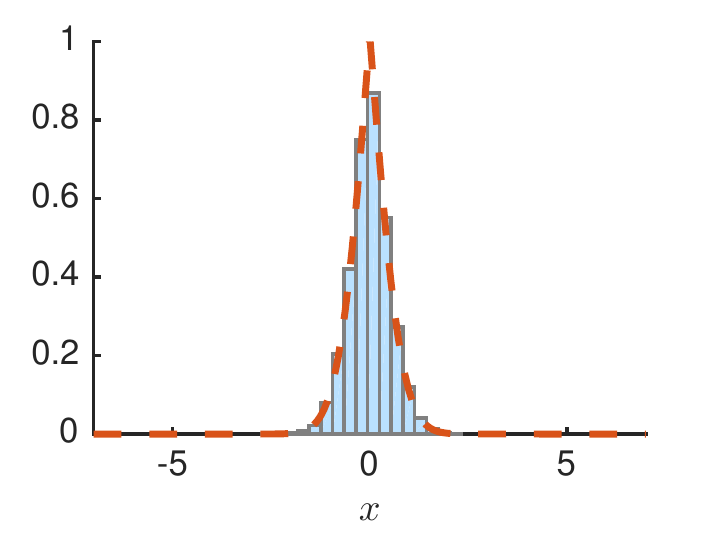}\includegraphics[scale=0.65]{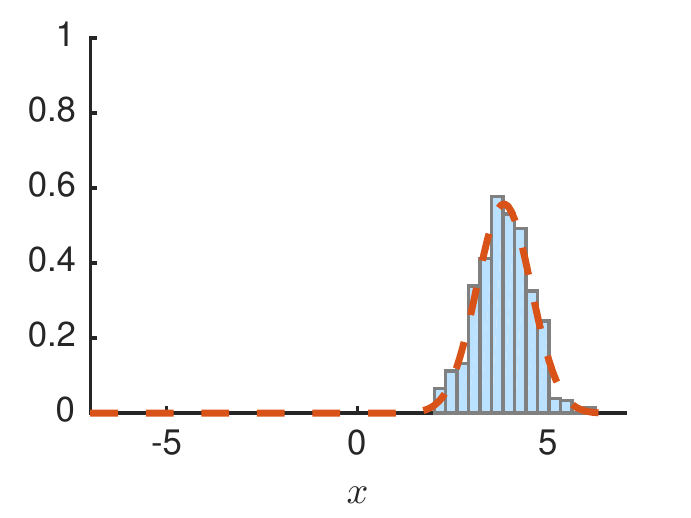}
	\vspace{-4ex}
	\caption{Theory v.s. simulations. The figures show comparisons between the limiting conditional densities $P_t(x\,\vert\,\xi)$ as predicted by the PDE \eref{strong_form} and the empirical densities obtained from Monte Carlo simulations. Top row: $t = 1$; bottom row: $t = 15$; left column: $\xi = 0$; and right column: $\xi = 1/\sqrt{\rho}$. See Example~\ref{ex:sim_PDE} for details of the experiment.}\label{fig:density}
\end{figure}

\begin{example}\label{ex:sim_PDE}
The proofs of Theorem~\ref{thm:limit} and Proposition~\ref{prop:q} will be presented elsewhere. Here, we verify the accuracy of the theoretical predictions made in them via numerical simulations. In our experiment, we generate a vector $\vxi$ whose components are i.i.d. and drawn from a marginal distribution $\pi(\xi)=(1-\rho)\delta(\xi) + \rho\delta(\xi-1/\sqrt{\rho})$. The sparsity level is set to $\rho = 0.05$. Starting from a random initial estimate $\vx_0$ with i.i.d. entries drawn from a normal distribution $\mathcal{N}(\frac{1}{\sqrt{2}}, \frac{1}{2})$, we use the OIST version of the online sparse PCA algorithm (see Example~\ref{ex:oist}) to estimate $\vxi$. The dimension is set to $p = 10,000$, and the other parameters are $\tau = 0.5, \beta = 0.27$, and $\omega = 1$. 

In \fref{density}, we compare the predicted limiting conditional densities $P_t(x\,\vert\,\xi = 0)$ and $P_t(x\,\vert\,\xi = 1/\sqrt{\rho})$ against the empirical densities observed in the simulations, at two different times ($t = 1$ and $t = 15$.) The PDE in \eref{strong_form} is solved numerically. We can see from the figure that the limiting densities given by the theory provide accurate predictions for the simulation results. In \fref{qt}, we verify the limiting form of the cosine overlap $Q_t$ as given in \eref{qt2}. For simulations, we average over $120$ independent instances of OIST, and plot the mean values and confidence intervals ($\pm 2$ standard deviations.) Again, we can see that the asymptotic results match with simulation data very well. Also shown in the figure are results for the standard Oja's method, for which we can obtain a \emph{closed-form} analytical formula for $Q_t$. This is the focus of the following subsection.
\end{example}

\subsection{The Nonsparse Case: Oja's Method}
\label{sec:Oja}

As mentioned earlier, the classical Oja's method \cite{Oja:1985} can be viewed as a special case of the algorithm in \eref{oist}. It corresponds to setting $\phi(x) = 0$ in \eref{eta}, \emph{i.e.}, the algorithm does not apply the nonlinear mapping $\eta(x)$. In this case, the limiting PDE \eref{strong_form} can be converted to a linear Fokker-Planck equation for the Ornstein-Uhlenbeck process, for which analytical solutions exist. For brevity, we omit discussions of this analytical solution of the PDE. Instead, we show a related result regarding the cosine similarity $Q_t$, which is an important figure of merit for the algorithm.

\begin{figure}[t]
	\centering{}
	\includegraphics[scale=0.7]{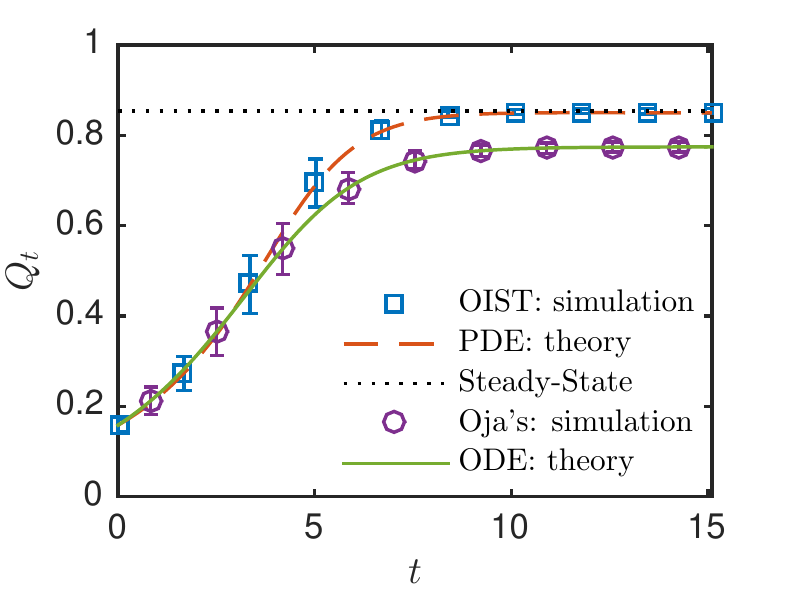}
	\vspace{-2ex}
  \caption{The comparison between the analytical predictions of the cosine similarity $Q_t$ and Monte Carlo simulations. For OIST, the theoretical curve is computed by using \eref{qt2}; for Oja's method, we use the closed-form formula in \eref{qt_Oja}. The theoretical predictions are plotted as dashed and solid lines, whereas the average values of $120$ Monte Carlo simulations are plotted as squares and circles. The error bars show confidence intervals of $\pm 2$ standard deviations. The black dotted line indicates the theoretical prediction of the steady-state given by the solution of the fixed-point equations in \eref{fp}.}\label{fig:qt}
\end{figure}

\begin{proposition}
For Oja's method, assume that we start the algorithm with a nonzero cosine similarity, \emph{i.e.}, $Q_0^p \rightarrow Q_0 \neq 0$ as $p \rightarrow \infty$. Then the dynamics of the cosine similarity $Q_t^p \rightarrow Q_t$, where $Q_t$ is given by
\begin{equation}\label{eq:qt_Oja}
Q_t^2 = \begin{cases}
    \alpha_{2} \left[\alpha_{1}+ \left( \frac{\alpha_{2}}{Q^{2}_0}-\alpha_{1} \right)e^{-2\alpha_{2}t} \right]^{-1} & \text{ if } \alpha_2\neq0 \\
    \left( 2 \alpha_1 t + Q_0^{-2} \right)^{-1} & \text{ if } \alpha_2=0.
  \end{cases}
\end{equation}
Here, $\alpha_1= \tau \omega(1+ \frac{\tau}{2})$ and $\alpha_{2}= \tau (\omega- \frac{ \tau}{2})$.
\end{proposition}
\begin{IEEEproof}[Proof (sketch)] We substitute $f(x, \xi) = x\xi$ into the weak form \eref{weak_form} of the limiting PDE. The left-hand side is then exactly $Q_t$. Using the facts that $\EE_\xi \xi^2 = 1$, $\phi(t) = 0$, and after some manipulations, we can simplify the right-hand side of \eref{weak_form} and get $Q_t = Q_0 + \int_0^t (-\alpha_1 Q_s^3 + \alpha_2 Q_s) \dif s$. Solving this ordinary differential equation leads to \eref{qt_Oja}.
\end{IEEEproof}

In the long-time limit, we have
\begin{equation}\label{eq:ode_steady}
	\lim_{t \to \infty}Q^{2}_t= \max \Bigg\{ 0, \frac{ \omega- \frac{ \tau}{2}}{ \omega(1+ \frac{ \tau}{2})} \Bigg\} . 
\end{equation}
This result indicates that for any finite step size $\tau>0$, Oja's method for online PCA cannot reach perfect estimation (\emph{i.e.}, $Q_\infty = 1$) even with infinite number of samples. Moreover, the formula also points out a simple phase transition phenomenon: when $\tau > 2 \omega$, the estimates obtained by the algorithm will be uncorrelated with $\vxi$.

\section{Steady State Analysis and Phase Transitions}
\label{sec:steady}

In this section, we study the long-time limit of OIST for sparse PCA. This steady-state analysis reveals an interesting phase transition phenomenon associated with OIST, which we also briefly discuss.

In the long-time limit, upon reaching the steady-state, the left-hand side of \eref{strong_form} becomes $0$. It follows that the steady-state density functions satisfy the equation
\begin{equation}\label{eq:steady_state}
\frac{\tau^2(1+\omega Q^2)}{2} \tpd{}{x}\!P(x\,\vert\,\xi) = \Gamma(x, \xi, Q, R) P(x\,\vert\,\xi),
\end{equation}
where $P(x\,\vert\,\xi)$, $Q$, $R$ are the steady-state versions of $P_t(x\,\vert\,\xi)$, $Q_t$ and $R_t$, respectively. Solving \eref{steady_state} and expanding $\Gamma$ according to its definition in \eref{Gamma}, we find the steady-state conditional density in the form of a Boltzmann distribution:
\begin{equation}\label{eq:density_steady}
	P(x\,\vert\,\xi)= \frac{1}{Z_\xi}\exp\left(- \frac{h(Q, R) x^2 +  \Phi(x) -  \tau \omega Q \xi x}{g(Q)}\right), 
\end{equation}
where $Z_\xi$ is the partition function,
\begin{equation}\label{eq:g_h}
\begin{aligned}
g(Q) &= \tau^2 (1+ \omega Q^2)/2\\
h(Q, R) &= \big(\tau \omega Q^2 - R + g(Q)\big)/2
\end{aligned}
\end{equation}
and $\Phi(x)$ is an antiderivative of $\phi(x)$. Note that $\Phi(x)$ can be any such antiderivative, since any constant added to $\Phi(x)$ will be absorbed into the normalization constant $Z_\xi$.

It is important to emphasize that \eref{density_steady} is only an \emph{implicit} definition of the steady-state distribution. This is because the expression relies on two constants $Q$ and $R$, whose values are determined by the self-consistent equations \eref{qt2} and \eref{rt2} (with $t \rightarrow \infty$) involving $P(x\,\vert\,\xi)$.

In what follows, we focus on OIST as discussed in Example~\ref{ex:oist}. Here, $\phi(x) = \beta \sgn(x)$, and thus we can set $\Phi(x) = \beta \abs{x}$. It follows that the exponent in \eref{density_steady} is a piecewise quadratic polynomial. This convenient form allows us to further simplify the right-hand sides of \eref{qt2} and \eref{rt2}. After some manipulations (which are omitted here), we can obtain the following fixed-point equations for determining $Q$ and $R$:
\begin{equation}\label{eq:fp}
\begin{aligned}
Q & = \sqrt{ \frac{g(Q)}{h(Q,R)}} \EE_{ \xi} \Big(\xi \frac{z_{+}f(z_{+})-z_{-}f(z_{-})}{f(z_{+})+f(z_{-})}\Big),                     \\
R & = \beta \sqrt{ \frac{g(Q)}{h(Q,R)}} \EE_{ \xi} \Big(\frac{ \frac{2}{ \pi}-z_{+}f(z_{+})-z_{-}f(z_{-})}{f(z_{+})+f(z_{-})}\Big), 
\end{aligned}
\end{equation}
where $g(Q), h(Q, R)$ are the functions defined in \eref{g_h}, $z_{ \pm}= \left(g(Q)h(Q,R) \right)^{- \frac{1}{2}} \left( \beta \pm \tau \w \xi Q   \right)/2$, and $f(\cdot)$ is the scaled complimentary error function defined as $f(x)= \frac{2}{ \pi}e^{x^{2}} \int_{x}^{ \infty}e^{-z^{2}}dz$. 

One can check that $\set{Q^0 = 0, R^0 = \frac{\tau^2}{2}}$ is always a solution to the fixed-point equations \eref{fp}. We call any such solution with $Q = 0$ an \emph{uninformative} solution, since it corresponds to a final estimate $\vx$ that is uncorrelated with $\vxi$. It is also revealing to examine the corresponding steady-state distributions. Substituting $Q^0, R^0$ into \eref{density_steady}, we find that, for any $\xi$, the conditional density is of the form
\begin{equation}\label{eq:Laplace}
P(x\,\vert\,\xi) = \frac{\beta}{\tau^2}e^{- \frac{2 \b}{ \tau^2} \abs{x}}.
\end{equation}
Since $P(x\,\vert\,\xi)$ does not depend on $\xi$, the variables $x$ and $\xi$ are independent; thus, the estimate provided by the algorithm contains no information about $\xi$ in the long-time limit.

In the low SNR regime, such uninformative fixed-points are the only solutions to \eref{fp}. The situation improves when we increase the SNR parameter $\omega$. At a certain critical value $\omega_c$, a nontrivial fixed point $\set{Q^\ast, R^\ast}$ with $Q^\ast \neq 0$ emerges. This corresponds to the case when the estimate $\vx$ becomes informative. We will present more detailed analysis of this phase transition phenomenon elsewhere. In what follows, we illustrate it using a numerical example.

We consider OIST at different SNR values. The other parameters in the algorithm are the same as those used in Example~\ref{ex:sim_PDE}. The left-side of \fref{steady} shows the limiting steady-state conditional densities $P(x\,\vert\,\xi = 1/\sqrt{\rho})$ for increasing values of the SNR parameter $\omega$. At a low SNR value ($\omega = 0.15$), we get the zero-mean (uninformative) Laplace distribution in \eref{Laplace}. As $\omega$ increases, the modes of the conditional densities move towards $1/\sqrt{\rho}$, starting to reveal information about $\xi$. In the right-side of \fref{steady}, we show the steady-state values of the cosine overlap $Q$ as a function of $\omega$. A clear phase transition appears at a critical value $\omega_c$. The theoretical prediction (the solid line in the figure), obtained by numerically solving the fixed-point equations \eref{fp}, matches very well with Monte Carlo simulations of the algorithm (shown as red dots.) Also shown in the figure are the results for Oja's method, with its theoretical prediction given by \eref{ode_steady}. Comparing OIST with Oja's method, we see that OIST has a lower phase transition threshold and that it also achieves a higher steady-state value for $Q$. This improvement in performance can be attributed to the fact that OIST exploits the sparsity structure of $\xi$ via iterative thresholding.

\begin{figure}[t]
	\centering
	\includegraphics[scale=0.66]{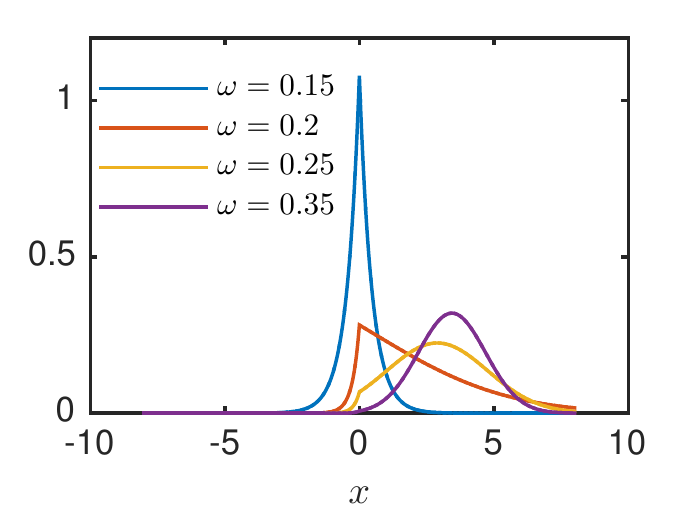}\includegraphics[scale=0.65]{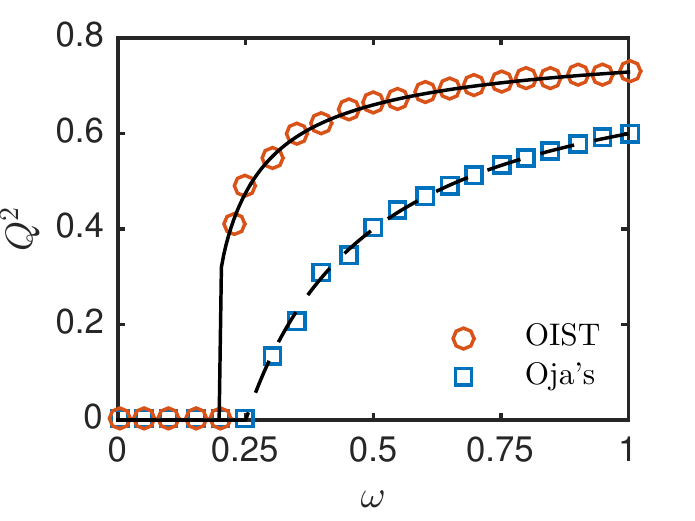}
	\vspace{-3ex}
	\caption{Steady-state distributions and phase transitions. Left-hand side: The steady-state densities $P(x\,\vert\,\xi=1/\sqrt{\rho})$ at different SNR values. Right-hand side: Theoretical predictions of the steady-state cosine overlap $Q$ as a function of the SNR parameter $\omega$. Black solid line: theoretical prediction for OIST; red dots: simulation results; black dashed line: theoretical prediction for Oja's method; blue squares: simulation results.}\label{fig:steady}
\end{figure}

\section{Conclusion}
\label{sec:conclusion}

We analyzed the dynamics of an online sparse PCA algorithm in the high-dimensional limit. The joint empirical measure of the underlying sparse eigenvector and its estimate as provided by the algorithm converges weakly to a deterministic process, characterized as the unique solution of a nonlinear PDE. This scaling limit provides exact information regarding the asymptotic performance of the algorithm. As a special case, we derived a closed-form expression for the limiting dynamics of the cosine similarity associated with Oja's method, a classical algorithm for online PCA. We also studied the steady-state of the nonlinear PDE and observed a phase transition phenomenon. The theoretical framework in this work is general. It paves the way towards understanding the dynamics of other online algorithms for various high-dimensional estimation problems. The theoretical analysis also provides insights and can lead to more principled ways of optimizing parameters in the algorithm to further improve performance.

\bibliographystyle{IEEEtran}

\end{document}

%% file: commands.tex
\usepackage{amsmath}
\usepackage{amssymb}
\usepackage{latexsym}
\usepackage{verbatim}
\usepackage{subfigure}
\usepackage[final]{graphicx}
\usepackage{psfrag}

\newtheorem{theorem}{Theorem}
\newtheorem{proposition}{Proposition}

\newtheorem{remark}{Remark}

\newtheorem{example}{Example}

{\begin{list}               
    {$\bullet$ \hfill}{
        \setlength{\leftmargin}{\parindent}
        \setlength{\parsep}{0.04\baselineskip}
        \setlength{\itemsep}{0.5\parsep}
        \setlength{\labelwidth}{\leftmargin}
        \setlength{\labelsep}{0em}}
    }
{\end{list}}

\providecommand{\eref}[1]{\eqref{eq:#1}}  
\providecommand{\cref}[1]{Chapter~\ref{chap:#1}}
\providecommand{\sref}[1]{Section~\ref{sec:#1}}
\providecommand{\fref}[1]{Figure~\ref{fig:#1}}

\providecommand{\R}{\ensuremath{\mathbb{R}}}

\providecommand{\abs}[1]{\lvert#1\rvert}
\providecommand{\norm}[1]{\lVert#1\rVert}
\providecommand{\inprod}[1]{\left\langle#1\right\rangle}
\providecommand{\set}[1]{\left\{#1\right\}}

\providecommand{\bydef}{\overset{\text{def}}{=}}

\renewcommand{\vec}[1]{\ensuremath{\boldsymbol{#1}}}
\providecommand{\mat}[1]{\ensuremath{\boldsymbol{#1}}}

\providecommand{\mI}{\mat{I}}

\providecommand{\va}{\vec{a}}

\providecommand{\vx}{\vec{x}} \providecommand{\vy}{\vec{y}}

\providecommand{\vxi}{\vec{\xi}}

\providecommand{\w}{\omega}

%% file: spca.bbl
\begin{thebibliography}{10}
\providecommand{\url}[1]{#1}
\csname url@samestyle\endcsname
\providecommand{\newblock}{\relax}
\providecommand{\bibinfo}[2]{#2}
\providecommand{\BIBentrySTDinterwordspacing}{\spaceskip=0pt\relax}
\providecommand{\BIBentryALTinterwordstretchfactor}{4}
\providecommand{\BIBentryALTinterwordspacing}{\spaceskip=\fontdimen2\font plus
\BIBentryALTinterwordstretchfactor\fontdimen3\font minus
  \fontdimen4\font\relax}
\providecommand{\BIBforeignlanguage}[2]{{%
\expandafter\ifx\csname l@#1\endcsname\relax
\typeout{** WARNING: IEEEtran.bst: No hyphenation pattern has been}%
\typeout{** loaded for the language `#1'. Using the pattern for}%
\typeout{** the default language instead.}%
\else
\language=\csname l@#1\endcsname
\fi
#2}}
\providecommand{\BIBdecl}{\relax}
\BIBdecl

\bibitem{Johnstone:2001}
I.~M. Johnstone, ``On the distribution of the largest eigenvalue in principal
  components analysis,'' \emph{Ann. Stat.}, vol.~29, no.~2, pp. 295--327, Apr.
  2001.

\bibitem{Nadler:2008}
B.~Nadler, ``{Finite sample approximation results for principal component
  analysis: A matrix perturbation approach},'' \emph{Ann. Stat.}, vol.~36,
  no.~6, pp. 2791--2817, 2008.

\bibitem{Johnstone:2009}
I.~M. Johnstone and A.~Y. Lu, ``On consistency and sparsity for principal
  components analysis in high dimensions,'' \emph{J. Am. Stat. Assoc.}, vol.
  104, no. 486, pp. 682--693, Jun. 2009.

\bibitem{Jolliffe:2003}
I.~T. Jolliffe, N.~T. Trendafilov, and M.~Uddin, ``A modified principal
  component technique based on the {LASSO},'' \emph{J. Comp. Graph. Stat.},
  vol.~12, no.~3, pp. 531--547, Sep. 2003.

\bibitem{Zou:2006}
H.~Zou, T.~Hastie, and R.~Tibshirani, ``Sparse principal component analysis,''
  \emph{J. Comp. Graph. Stat.}, vol.~15, no.~2, pp. 265--286, 2006.

\bibitem{dAspremont:2007}
A.~d'Aspremont, L.~El~Ghaoui, M.~I. Jordan, and G.~R.~G. Lanckriet, ``A direct
  formulation for sparse {PCA} using semidefinite programming,'' \emph{{SIAM}
  Rev.}, vol.~49, no.~3, pp. 434--448, Jan. 2007.

\bibitem{Amini:2009}
A.~A. Amini and M.~J. Wainwright, ``High-dimensional analysis of semidefinite
  relaxations for sparse principal components,'' \emph{Ann. Stat.}, vol.~37,
  no.~5B, pp. 2877--2921, Oct. 2009.

\bibitem{Shen:2008}
H.~Shen and J.~Z. Huang, ``Sparse principal component analysis via regularized
  low rank matrix approximation,'' \emph{J. Maultivar. Anal.}, vol.~99, no.~6,
  pp. 1015--1034, Jul. 2008.

\bibitem{Journee:2008}
M.~Journ{\'e}e, Y.~Nesterov, P.~Richt{\'a}rik, and R.~Sepulchre, ``Generalized
  power method for sparse principal component analysis,'' \emph{J. Mach. Learn.
  Res.}, vol.~11, no. 517--553, 2010.

\bibitem{Ma:2013}
Z.~Ma, ``Sparse principal component analysis and iterative thresholding,''
  \emph{Ann. Stat.}, vol.~41, no.~2, pp. 772--801, Apr. 2013.

\bibitem{Deshpande:2014}
Y.~Deshpande and A.~Montanari, ``Information-theoretically optimal sparse
  {PCA},'' in \emph{{IEEE} {International} {Symposium} on {Information}
  {Theory}}, 2014.

\bibitem{Mairal:2009}
J.~Mairal, F.~Bach, J.~Ponce, and G.~Sapiro, ``Online learning for matrix
  factorization and sparse coding,'' \emph{J. Mach. Learn. Res.}, vol.~11, pp.
  19--60, 2010.

\bibitem{Arora:2012}
R.~Arora, A.~Cotter, K.~Livescu, and N.~Srebro, ``Stochastic optimization for
  {PCA} and {PLS},'' in \emph{Proc. 50th {Annual} {Allerton} {Conference} on
  {Communication}, {Control}, and {Computing} ({Allerton})}, Oct. 2012.

\bibitem{Balsubramani:2013}
A.~Balsubramani, S.~Dasgupta, and Y.~Freund, ``{The fast convergence of
  incremental PCA},'' in \emph{Adv. Neural Inf. Process. Syst.}, 2013.

\bibitem{Yang:2015}
W.~Yang and H.~Xu, ``Streaming sparse principal component analysis,'' in
  \emph{Proceedings of the 32nd {International} {Conference} on {Machine}
  {Learning} ({ICML}-15)}, 2015, pp. 494--503.

\bibitem{Oja:1985}
E.~Oja and J.~Karhunen, ``On stochastic approximation of the eigenvectors and
  eigenvalues of the expectation of a random matrix,'' \emph{J. Math. Anal.
  Appl.}, vol. 106, no.~1, pp. 69--84, 1985.

\bibitem{Parikh:2014}
N.~Parikh and S.~Boyd, ``Proximal {Algorithms},'' \emph{Foundations and Trends
  in Optimization}, vol.~1, no.~3, Jan. 2014.

\bibitem{Billingsley:1999}
P.~Billingsley, \emph{Convergence of probability measures}, 2nd~ed.\hskip 1em
  plus 0.5em minus 0.4em\relax New York: Wiley, 1999.

\bibitem{Kallenberg:2002}
O.~Kallenberg, \emph{Foundations of modern probability}, 2nd~ed.\hskip 1em plus
  0.5em minus 0.4em\relax Springer, 2002.

\bibitem{Risken:1996}
H.~Risken, \emph{The {Fokker}-{Planck} equation: {M}ethods of solution and
  applications}, 2nd~ed.\hskip 1em plus 0.5em minus 0.4em\relax New York:
  Springer-Verlag, 1996.

\end{thebibliography}
